\pdfoutput=1
\documentclass[a4paper]{article}

\usepackage{INTERSPEECH2022}
\usepackage{hyperref}
\usepackage{graphicx}
\usepackage{multirow,multicol, makecell}
\let\svthefootnote\thefootnote
\newcommand\blankfootnote[1]{%
  \let\thefootnote\relax\footnotetext{#1}%
  \let\thefootnote\svthefootnote%
}

\title{ECAPA-TDNN for Multi-speaker Text-to-speech Synthesis}
\name{
    Jinlong Xue$^1$, 
    Yayue Deng$^{1,2}$,
    Yichen Han$^1$,
    Ya Li$^{1,*}$,
    Jianqing Sun$^3$,
    Jiaen Liang$^3$
}
\address{
  $^1$School of Artificial Intelligence, Beijing University of Posts and Telecommunications,\\ Beijing, China\\
  $^2$Beijing Language and Culture University, Beijing, China \\
  $^3$Unisound AI Technology Co., Ltd, Beijing, China}
  
\email{
    jinlong\_xue@bupt.edu.cn,
    dengyayue.stu@gmail.com,
    adelacvgaoiro@bupt.edu.cn,
    yli01@bupt.edu.cn,
    sunjianqing@unisound.com,
    liangjiaen@unisound.com
}

\begin{document}

\maketitle
\blankfootnote{* Ya Li is the corresponding author.}

\begin{abstract}

    In recent years, neural network based methods for multi-speaker text-to-speech synthesis (TTS) have made significant progress. However, the current speaker encoder models used in these methods still cannot capture enough speaker information. In this paper,  we focus on accurate speaker encoder modeling and propose an end-to-end method that can generate high-quality speech and better similarity for both seen and unseen speakers. The proposed architecture consists of three separately trained components: a speaker encoder based on the state-of-the-art ECAPA-TDNN model which is derived from speaker verification task, a FastSpeech2 based synthesizer, and a HiFi-GAN vocoder. The comparison among different speaker encoder models shows our proposed method can achieve better naturalness and similarity. To efficiently evaluate our synthesized speech, we are the first to adopt deep learning based automatic MOS evaluation methods to assess our results, and these methods show great potential in automatic speech quality assessment.
    
\end{abstract}

\section{Introduction}

    Text-to-speech (TTS) aims to produce natural human speech. In the past few years, deep learning based models have developed rapidly. Recent research shows that the quality and the naturalness of the synthesized voices are comparable with real human speech, such as Tacotron 2~\cite{shen2018natural}, DeepVoice 3~\cite{ping2017deepvoice3}, and FastSpeech 2~\cite{ren2020fastspeech}. Despite the successful achievement of speaker-dependent TTS, how to create expressive and controllable in terms of various speaking styles in multi-speaker task still needs more research. On the other hand, the models of the few-shot voice cloning in unseen speakers circumstance by using a speaker encoder usually tend to synthesize neutral and poor quality voices compared to the real speaker. Therefore, how to sufficiently extract speaker information from the reference voices becomes significant.
    
    To accomplish the multi-speaker task, a TTS system and a speaker representation are needed. In previous studies, most multi-speaker systems use a speaker encoder to extract speaker embedding to characterize the target speaker's voice and style. Because models in speaker verification task are designed to extract the text-independent speaker embeddings from the target speaker voices to capture speaker characteristics, they have been widely adopted as the speaker encoder, such as d-vector~\cite{variani2014deep}, x-vector~\cite{snyder2018x}. Besides, pretrained models are more often used instead of jointly training with the TTS system, for the speaker knowledge for speaker encoder is limited by training dataset in the latter case. Jia et al.~\cite{jia2018transfer} investigated the knowledge transfer where the speaker verification model is trained on a dataset with many speakers, like VoxCeleb~\cite{Nagrani17VoxCeleb, Chung18bVoxCeleb2} dataset. Thus, the speaker embedding extracted from the speaker verification model conditioning the TTS system leads to better generalization and performance on the multi-speaker TTS and the voice cloning task. Especially the combination of x-vector~\cite{snyder2018x} and TTS system achieves promising results~\cite{cooper2020zero}.
    
    However, the naturalness and speaker similarity of the audios synthesized from the current models are less favorable, especially in unseen datasets. The reason is that the ability of the current speaker encoders is not enough to capture enough information of the target speakers in the multi-speaker TTS task. To address these weaknesses, we propose our multi-speaker TTS by adopting the non-autoregressive TTS model FastSpeech 2 and the TDNN-based model ECAPA-TDNN~\cite{desplanques2020ecapa} from speaker verification task, which has stronger speaker features extraction ability and robustness~\cite{dawalatabad2021ecaparobust}. It introduces multiple enhancements to the basic architecture and outperforms other TDNN based speaker verification models on the VoxCeleb datasets. We compare different speaker encoders and investigate their generalization ability in two publicly available datasets for both the seen and unseen tests. Our method outstands other methods in both naturalness and speaker similarity.
    
    To better evaluate our methods, we need many subjective evaluations including the mean opinion score (MOS) test and speaker similarity test. However, such measurement requires many humans to be involved, making it time-consuming and expensive. Thanks to the VCC 2016 and VCC 2018 datasets~\cite{toda2016voice,lorenzo2018voice}, several deep-learning-based automatic speech quality evaluation methods have been proposed, such as MOSNet~\cite{lo2019mosnet}, MBNet~\cite{leng2021mbnet}, a self-supervised representation based MOS predictor model (denoted as S3PRL)\cite{tseng2021utilizing}. To our best knowledge, we are the first to evaluate the synthesized speech by using the MOS prediction model to accelerate our research. We utilize these methods and compare the MOS score results. The results obtained from automatic MOS prediction models are consistent with subjective MOS results, which shows that they have great potential to free us from the burden of MOS tests.
    
    The paper is organized as follows: Section~\ref{sec:model} describes related works in terms of speaker representations, and Section~\ref{sec:proposed} illustrates our proposed method with training workflow. Experimental setup and results are shown in Section~\ref{sec:exp}. At last, we conclude our finding in Section~\ref{sec:conclusion}. Examples of synthesized speech can be found on the project page\footnote{Audio samples: https://happylittlecat2333.github.io/interspeech2022}.

\begin{figure*}[t]
  \centering
  \includegraphics[width=\linewidth]{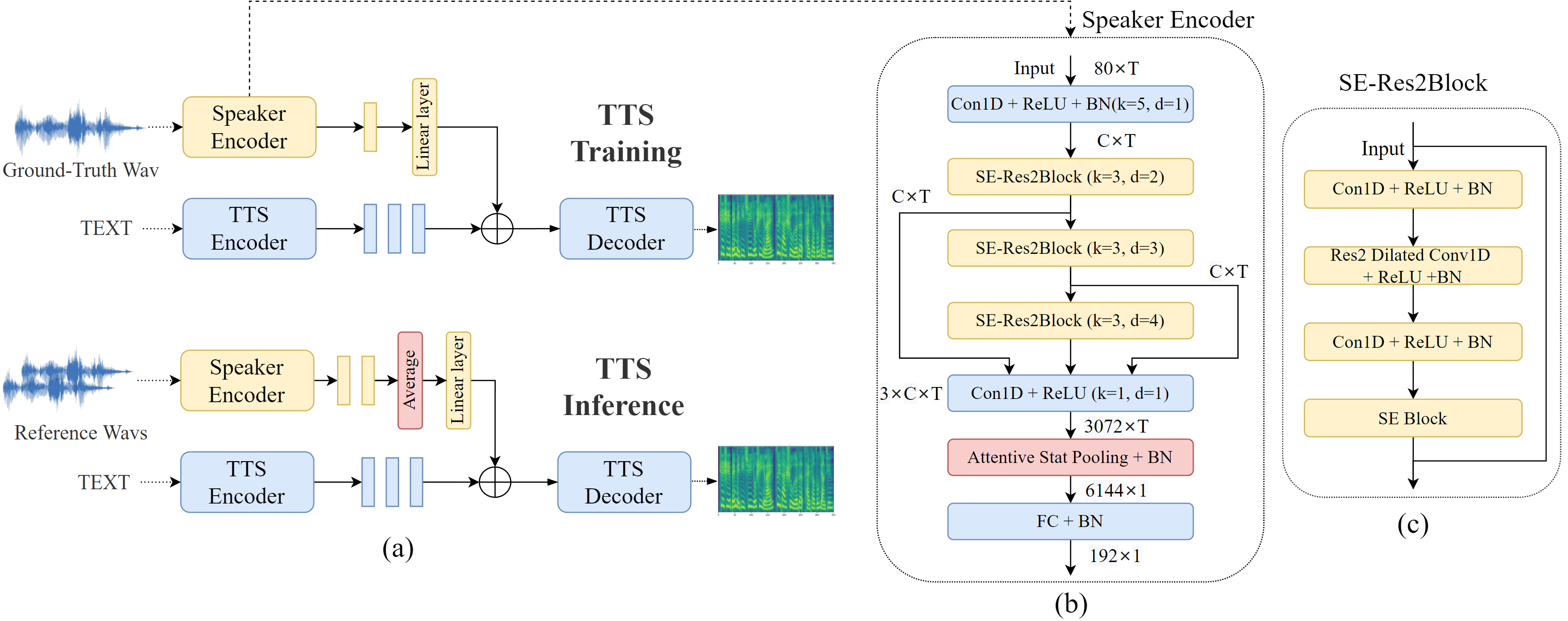}
  \caption{
  The illustration of our proposed ECAPA-TDNN based multi-speaker speech synthesis model. (a) depicts the training and inference workflows in our experiments. (b) shows the whole network of ECAPA-TDNN. K stands for kernel size and d for dilation. C and T denote channels and temporal dimensions. (c) is the detailed illustration of SE-Res2Block in the ECAPA-TDNN model.
  }
  \label{fig:model_illustration}
\end{figure*}

\section{Speaker Representations}
\label{sec:model}


    Multi-speaker TTS system highly depends on speaker representation for conditioning the acoustic model to clone the voices from target speakers. Many models derived from the speaker verification task have been applied in TTS system because they can extract speaker information from speech. These models are usually trained on a large number of text-independent datasets~\cite{Nagrani17VoxCeleb, Chung18bVoxCeleb2} recorded by many speakers, which provides them the ability to capture the subtle characteristics and styles from different speakers by using only a short utterance under any circumstances.
    
    In speaker verification task, deep neural network models have surpassed the classic model i-vector~\cite{ibrahim2018ivector}. Among them, d-vector~\cite{variani2014deep}, x-vector~\cite{snyder2018x} are representative methods, and they have been used in multi-speaker TTS~\cite{jia2018transfer, cooper2020zero}. These combinations show great potential and some methods have been widely used~\cite{snyder2018x}. Below, we describe each of these speaker encoder models.

\subsection{D-vector}

    The conventional d-vector~\cite{variani2014deep} model uses a DNN architecture as a speaker feature extractor operating at the frame level.
    After sending the frames through the DNN network, d-vector is obtained by element-wise averaging the frame-level 
    outputs from the last hidden layer of DNN network.
    
\subsection{X-vector}

    In the original x-vector paper~\cite{snyder2018x}, a time-delay neural network (TDNN) with sub-sampling is used as the encoder network. An attentive statistics pooling (ASP)~\cite{okabe2018attentive} layer aggregates all frame-level outputs from the last encoder layer and computes its mean and standard deviation. After sending through segment-level layer (fully-connected layer), the x-vector speaker embedding is obtained.
    The combination of a TDNN network with dilation has three advantages: reduction of the total number of independent connection parameters, invariance under shifts in times, and larger contextual vision. The use of ASP allows the model to select frames that are indeed relevant to speaker characteristics.

\section{Proposed Method}
\label{sec:proposed}

    Inspired by the outstanding performance of ECAPA-TDNN in speaker verification task, we introduce a speaker encoder based on this model to our multi-speaker TTS system. Fig.~\ref{fig:model_illustration} shows our modified speaker encoder based on ECAPA-TDNN and our proposed method with training and inference workflows.
    
\subsection{Speaker encoder}

    Based on recent trends in the related field of computer vision and face verification, ECAPA-TDNN also uses TDNN as its base architecture but introduces multiple enhancements: using Squeeze-and-Excitation (SE) blocks~\cite{hu2018squeeze} in encoder modules to explicitly model channel interdependencies; implementing Res2Net with skip connections;  aggregating and propagating features of different hierarchical levels in the encoder to capture both the shallow and deep speaker feature maps; improving statistics pooling module with channel- and context-dependent frame attention to focus more on speaker-specific characteristics such as focusing more vowels than consonants. Those improvements endow ECAPA-TDNN with the ability to extract subtle speaker information and outperform other TDNN based models.
    
\subsection{Acoustic model}

    We extend the non-autoregressive model FastSpeech 2~\cite{ren2020fastspeech} architecture to implement our multi-speaker model. FastSpeech 2 is composed of Transformer-based encoder and decoder with a variant adaptor. The encoder generates the hidden embedding from a sequence of phoneme-level inputs. The variant adaptor aims to add variant information to phoneme hidden sequences and it is composed of a duration predictor, a pitch predictor, and an energy predictor. Finally, the decoder generates the mel spectrogram from the hidden sequences expanded by the variant adaptor. Following the module in Tacotron~\cite{wang2017tacotron},  we add a Postnet (Conv1D blocks) module after the decoder to finetune the speech quality.


    
\subsection{TTS training and inference}

    At the training stage, we use the speaker encoder model pretrained on speaker verification task to extract fixed-dimensional embeddings from each utterance for speaker representations. After that, the utterance-level speaker representations are projected to match the dimension of the output from the encoder of acoustic model with one linear layer, and they are expanded and added to the output. Therefore, speaker information is transferred to the synthesizer and the variant adaptor in FastSpeech 2 can be conditioned on speaker information. At the inference stage, we extract a speaker representation for each utterance and compute an average representation on behalf of this speaker. The other process is the same as the training stage.
    


\section{Experiments}
\label{sec:exp}

\subsection{Experimental Setup}

    We use two publicly available English datasets: VCTK~\cite{vctk} and LibriTTS~\cite{zen2019libritts}. VCTK corpus includes speech data uttered by $109$ English speakers with various accents at $48$ kHz. Each speaker reads out about $400$ sentences and about $44$ hours of data in sum. LibriTTS consists of $585$ hours of speech data at the $24$ kHz sampling rate from $2,456$ speakers and the corresponding texts. In our experiments, all utterances are down-sampled to $22050$ Hz and are used to extract $80$ dimensional mel spectrograms. 
    
    We implement pretrained x-vector\footnote{https://github.com/manojpamk/pytorch\_xvectors}, ECAPA-TDNN\footnote{https://huggingface.co/speechbrain/spkrec-ecapa-voxceleb} as speaker encoder. The x-vector and ECAPA-TDNN are both pretrained on Voxceleb 1 and Voxceleb 2~\cite{Nagrani17VoxCeleb, Chung18bVoxCeleb2}, and the EER results on Vox1-test are $2.82\%$ and $0.69\%$. For preprocessing, each utterance is resampled to $16$ kHz but converted to $30$ dimensional MFCC features for x-vector and $80$ for ECAPA-TDNN. After extraction, the dimensions of the speaker embeddings for x-vector and ECAPA-TDNN are $512$ and $128$ respectively.
    
    For the acoustic model, we follow the open-source FastSpeech2 implementation\footnote{https://github.com/ming024/FastSpeech2}. We use Montreal Force Aligner~\cite{mcauliffe2017montreal} to get the ground-truth duration for each phoneme as additional inputs.
    We use the pretrained HiFi-GAN\footnote{https://github.com/jik876/hifi-gan} model as our vocoder to convert the $80$ dimensional mel spectrograms to $22050$ Hz audio files. Our multi-speaker models are all trained for $400k$ steps with a batch size of $16$ on one GeForce RTX 3090. 
    
    We use the following methods for evaluation:
    \begin{enumerate}
    \item ground-truth: the real utterances from the datasets.
    \item reconstruct: directly convert the ground-truth mel spectrograms back to speech.
    \item baseline: FastSpeech 2 with look-up table.
    \item x-vector: FastSpeech 2 with pretrained x-vector speaker encoder.
    \item ecapa: our proposed method by combining FastSpeech 2 with pretrained ECAPA-TDNN speaker encoder.
    \end{enumerate}
    
    To comprehensively evaluate our proposed model, we split the VCTK dataset for training and testing: $8$ speakers are held as unseen speakers cloning test, and other $101$ speakers are used to train and evaluate models for seen speakers. We use LibriTTS for the unseen speaker cloning test. 
    
    During testing, we use the average speaker embeddings extracted from all the utterances of the same speaker instead of from only one utterance, because using the averaged embedding can be more stable and have better similarity in our experiment. It should also be noted that the pretrained models will not be finetuned and be adapted in useen speakers in our experiments to evaluate the voice cloing ability of our proposed method.

\subsection{Objective similarity evaluation}

    We use a third-party pretrained speaker encoder
    to evaluate the speaker similarity between the real speech and the synthesized speech. 
    To evaluate how similar synthesized speech and real speech are, we make pairs for each synthesized utterance with a randomly selected real utterance from the same speaker. Then we use the pretrained speaker encoder to extract speaker embeddings for each utterance and compute the average cosine similarity for each pair as our similarity result.
    
    The results of the objective speaker similarity test are shown in Table~\ref{tab:objective_similarity}. It can be seen that using the pretrained ECAPA-TDNN model as speaker encoder outperforms x-vector in both seen speaker test and unseen speaker test in VCTK, even in unseen speaker test set from LibriTTS. Moreover, the proposed model matches the baseline (usually having best results in seen speaker) in seen speaker test on VCTK. For unseen speaker test on VCTK and LibriTTS, using a simple lookup table cannot clone unseen speaker voices, and our proposed method has better similarity than the x-vector model, which suggests ECAPA-TDNN model can capture more speaker information and have potential to utilize in multi-speaker task than other speaker encoder models.

\subsection{Subjective evaluation}

    We conduct a mean opinion score (MOS) test to evaluate the naturalness and speaker similarity of the synthesized speech. In our test, we randomly select 20 utterances from each test set. In the quality MOS test, the listeners are given one synthesized utterance and are asked to give a rating score between 1 to 5 points for speech quality. In the speaker similarity test, the listeners are given both a real utterance and a synthesized utterance to evaluate the similarity by scoring between 1 to 5 points. Table~\ref{tab:subjective} shows the MOS results and the speaker similarity results.
    
    In naturalness test, it can be seen that our proposed method outperforms the x-vector model in all test sets including unseen speaker tests. In similarity test, it shows that the speech synthesized by our model also has better similarity than other methods. The results are consistent with the objective similarity evaluation. These results suggest that combining the ECAPA-TDNN model and acoustic models has the power to gain better speech naturalness and speaker similarity in the multi-speaker task.

\subsection{Automatic MOS evaluation}

    To further evaluate the effectiveness of our proposed method, we use several automatic speech quality assessment models to assess our synthesized speech.
    We use three pretrained MOS prediction models in our experiment: MOSNet, MBNet, and Self-supervised Representation method (S3PRL). These models are pretrained on VCC 2016~\cite{toda2016voice} and VCC 2018~\cite{lorenzo2018voice} datasets, which include lots of MOS rating scores evaluated by many participants.
    It should be noticed that the speech quality in the voice conversion task is less natural than speech synthesis task, and the rating scores in the VCC datasets are lower than the usual TTS score.
    We use the same MOS test set in subjective evaluation and the results of the objective MOS prediction test are shown in Table~\ref{tab:mos_predictor}.
    
    It can be seen from the results that the proposed method outperforms the x-vector model and achieves comparable speech quality to the baseline in seen VCTK test, which is consistent with the results in subjective MOS evaluation. Besides, our proposed model also outperforms the x-vector model in unseen VCTK or unseen LibriTTS test.

\begin{table}[t]
  \caption{Objective speaker similarity for different test sets and different types of speaker encoders.}
  \label{tab:objective_similarity}
  \centering
  \scalebox{1}{
      \begin{tabular}{c c c c c}
        \toprule
        \textbf{method} & 
        \makecell{\textbf{Seen} \\ \textbf{VCTK}} &
        \makecell{\textbf{Unseen} \\ \textbf{VCTK}} &
        \makecell{\textbf{Unseen} \\ \textbf{LibriTTS}} \\
        \midrule
        {ground-truth}& .979          & .975              & .986              \\
        {reconstruct} & .976          & .972              & .984              \\
        {baseline}    & \textbf{.968} & -                 & -                 \\
        {x-vector}    & .963          & .954              & .956              \\
        {ecapa}       & .967          & \textbf{.959}     & \textbf{.959}     \\
        \bottomrule
      \end{tabular}
  }
\end{table}

\begin{table}[t]
  \caption{The results of the subjective MOS tests for naturalness and speaker similarity.}
  \label{tab:subjective}
  \centering
  \scalebox{0.96}{
      \begin{tabular}{c c c c c}
        \toprule
        \textbf{Model} & 
        \textbf{method} & 
        \makecell{\textbf{Seen} \\ \textbf{VCTK}} &
        \makecell{\textbf{Unseen} \\ \textbf{VCTK}} &
        \makecell{\textbf{Unseen} \\ \textbf{LibriTTS}} \\
        \midrule \midrule
        \multirow{5}{*}{MOS}        & ground-truth & 4.19          & 4.20          & 4.21  \\
                                    & reconstruct  & 4.09          & 4.11          & 4.10  \\
                                    & baseline     & \textbf{3.70} & -             & -     \\
                                    & x-vector     & 3.51          & 3.51          & 3.38  \\
                                    & ecapa        & 3.62          & \textbf{3.62} & \textbf{3.47} \\
        \midrule
        \multirow{5}{*}{Similarity} & reconstruct  & 4.65          & 4.69          & 4.66  \\
                                    & baseline     & 3.89          & -             & -     \\
                                    & x-vector     & 3.71          & 3.65          & 3.08  \\
                                    & ecapa        & \textbf{3.93} & \textbf{3.66} & \textbf{3.18} \\
        \bottomrule
      \end{tabular}
  }
\end{table}

\begin{table}[t]
  \caption{Automatic MOS evaluation results for seen and unseen test sets by using three MOS prediction models.}
  \label{tab:mos_predictor}
  \centering
  \scalebox{0.97}{
      \begin{tabular}{ccccc}
        \toprule
        \textbf{Model} & 
        \textbf{method} & 
        \makecell{\textbf{Seen} \\ \textbf{VCTK}} &
        \makecell{\textbf{Unseen} \\ \textbf{VCTK}} &
        \makecell{\textbf{Unseen} \\ \textbf{LibriTTS}} \\
        \midrule \midrule
        \multirow{5}{*}{MOSNet} & ground-truth & 4.16          & 3.91           & 3.40            \\
                                & reconstruct  & 3.75          & 3.83           & 3.34            \\
                                & baseline     & \textbf{3.32} & -              & -               \\
                                & x-vector     & 3.11          & 3.44           & 3.15            \\
                                & ecapa        & 3.16          & \textbf{3.52}  & \textbf{3.42}   \\
        \midrule
        \multirow{5}{*}{MBNet}  & ground-truth & 3.86          & 3.99           & 3.05            \\
                                & reconstruct  & 3.45          & 3.81           & 2.99            \\
                                & baseline     & \textbf{3.37} & -              & -               \\
                                & x-vector     & 3.07          & 3.46           & 3.21            \\
                                & ecapa        & 3.35          & \textbf{3.55}  & \textbf{3.53}   \\
        \midrule
        \multirow{5}{*}{S3PRL}  & ground-truth & 3.53          & 3.53          & 3.45            \\
                                & reconstruct  & 3.44          & 3.47          & 3.37            \\
                                & baseline     & \textbf{3.45} & -             & -               \\
                                & x-vector     & 3.27          & 3.42          & 3.36            \\
                                & ecapa        & 3.44          & \textbf{3.52} & \textbf{3.48}   \\
        \bottomrule
      \end{tabular}
  }
\end{table}

\subsection{Analysis}

\begin{figure}[t]
  \centering
  \includegraphics[width=0.85\linewidth]{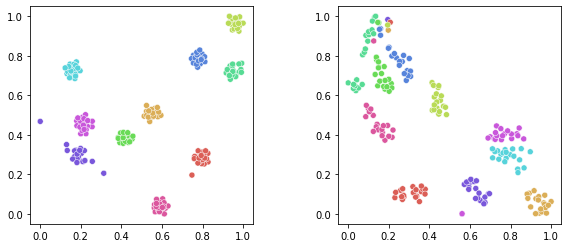}
  \caption{Visualizations of different speaker embeddings. \\\centering{Left: x-vector	Right: ECAPA-TDNN}}
  \label{fig:visualization}
\end{figure}

     In order to investigate why our proposed method has better performance in the multi-speaker TTS task, we visualize the speaker embeddings in Fig~\ref{fig:visualization}. We randomly select the speaker embeddings extracted from 200 utterances from 10 speakers and use t-SNE to reduce them into 2-dimension. It can be seen from the plot that 
     both the ECAPA-TDNN model and the x-vector model can discern the utterance from the same speaker, while the distribution of ECAPA-TDNN is more continuous which suggests that it 
     clusters each speaker but keeping the subtle speaker characteristics from different utterances spoken by the same speaker. This is helpful in multi-speaker synthesis, as its goal is different from speaker verification task.
     Previous studies~\cite{chien2021investigating} suggest that the continuous distribution of speaker embeddings has better performance in the multi-speaker TTS task.
     Our experiment results in similarity tests confirm these studies~\cite{chien2021investigating}.
     As a result, using ECAPA-TDNN as a speaker encoder can achieve better speech naturalness and speaker similarity.
     
    By analyzing the results obtained from humans and MOS predictors, we find that assessments from automatic MOS predictors are consistent with evaluations from subjective method, which can both reflect the quality of the synthesized speech in the seen and unseen VCTK test set. The use of these models can free us from the burden of collecting subjective evaluations. Meanwhile, we also see that the currently MOS prediction models lose their effectiveness in some circumstances like unseen LibriTTS, and a more comprehensive MOS dataset may increase their robustness and accuracy.

\section{Conclusions}
\label{sec:conclusion}

    In order to improve the naturalness and speaker similarity in multi-speaker text-to-speech synthesis, we propose our end-to-end method by introducing a more powerful speaker encoder based on the ECAPA-TDNN model derived from speaker verification task. We combine the independently pretrained ECAPA-TDNN model with a non-autoregressive acoustic model FastSpeech2.
    By transferring the knowledge learned from other datasets and applying the SOTA speaker verification model, our proposed model outperforms other methods in both speech naturalness and speaker similarity. Besides, to lighten the burden of subjective evaluation, we are the first to adopt automatic MOS predictors to assess our testing results and these models show great potential. 
    For future work, we will continue to investigate the performance of few-shot multi-speaker speech synthesis.
    
\section{Acknowledgements} 
\label{sec:ack}

    This work is supported by the Open Project Program of the National Laboratory of Pattern Recognition (NLPR) (202200042) and New Talent Project of Beijing University of Posts and Telecommunications (2021RC37).

\bibliographystyle{IEEEtran}

\bibliography{mybib}


\end{document}